\documentclass{egpubl}
\usepackage{pg2026s}
\WsConferencePaper

\usepackage[T1]{fontenc}
\usepackage{dfadobe}

\usepackage{amsmath}
\usepackage{amssymb}
\usepackage{float}
\usepackage{cite}
\BibtexOrBiblatex
\electronicVersion
\PrintedOrElectronic
\ifpdf \usepackage[pdftex]{graphicx} \pdfcompresslevel=9
\else \usepackage[dvips]{graphicx} \fi

\usepackage{egweblnk}
\usepackage{microtype}
\usepackage{wrapfig}
\usepackage{multirow}
\usepackage{booktabs}

\title[Inverse Rig Optimization from Line Drawings]{Inverse Rig Optimization from Line Drawings}

\author[Z. Zhu \& Y. Koyama]
{\parbox{\textwidth}{\centering
Zihao Zhu$^{1}$\orcid{0009-0008-7980-8198}
and Yuki Koyama$^{1}$\orcid{0000-0002-3978-1444}
}
\\
{\parbox{\textwidth}{\centering
$^1$The University of Tokyo, Japan
}}
}

\begin{document}

\teaser{
  \centering
  \includegraphics[width=\linewidth]{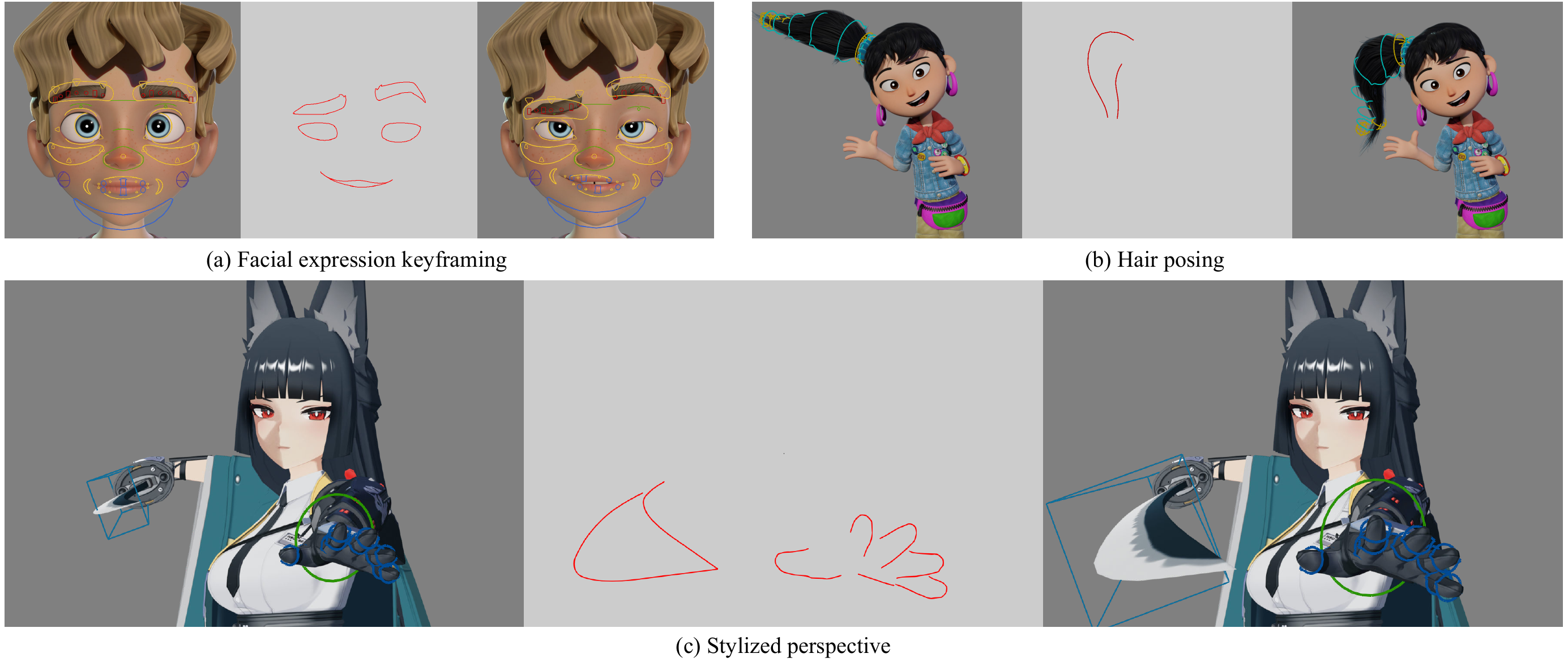}
  \caption{Given a rigged mesh, our method recovers the rig
  parameters that produce the artist's strokes. We show
  three applications. Each example consists of three panels: the
  initial pose with the rig controls overlaid (left), the
  artist's stroke input (middle), and the optimized pose
  produced by our method (right). Character models in (a) and (b)
  \textcopyright~Blender Studio; character model in (c) \textcopyright~miHoYo.}
  \label{fig:teaser}
}

\maketitle

\begin{abstract}
Stylized 3D character animation is largely hand-authored, with animators authoring rig parameters one keyframe at a time to find the best pose.
Because stylized work reads chiefly through contour lines, drawing contours in the camera view is the most direct and precise way to express artistic intent.
This mismatch between the rig controls and the artist's goal forces a laborious trial-and-error workflow, with animators repeatedly manipulating rig controls against the rendered view to match the desired contour.
To address this, we propose a method that recovers rig parameters from screen-space contour strokes, enabling effective keyframing from sketches.
Given strokes that redraw the current contour, our method optimizes the high-level rig parameters defined in the DCC tool.
The key is to use a pre-trained MLP rig surrogate that provides a differentiable map from rig parameters to mesh vertices, replacing the original black-box rig within the optimization process.
We match user-drawn lines to mesh contour lines and backpropagate the resulting screen-space error through the surrogate to update the rig parameters.
Our results demonstrate that the method works for diverse characters and practical scenarios.

\begin{CCSXML}
<ccs2012>
<concept>
<concept_id>10010147.10010371.10010352.10010379</concept_id>
<concept_desc>Computing methodologies~Animation</concept_desc>
<concept_significance>300</concept_significance>
</concept>
</ccs2012>
\end{CCSXML}

\ccsdesc[300]{Computing methodologies~Animation}

\printccsdesc
\end{abstract}

\section{Introduction}
\label{sec:intro}

Stylized 3D animation, spanning cartoon and anime styles, has become
increasingly popular as a mode of character animation, prized for an
expressive visual identity that photorealistic rendering cannot reproduce.
Its \emph{characters} depart from real-human anatomy by design: proportions
are exaggerated, faces are abstracted into a small number of iconic shapes,
and the body parts follow conventions drawn from 2D illustration rather
than from human anatomy. Likewise, the \emph{animation} that brings these
characters to life is equally non-physical, featuring anticipation-driven
squash and stretch, extreme facial poses, and view-dependent geometry
sculpted to read correctly from one specific camera. Each keyframe is
composed for a particular look on screen rather than for physical
plausibility.

Achieving these stylized visual effects is slow, manual, trial-and-error handwork. Because the
style is deliberately non-physical, it cannot be captured or
simulated, so the animator must author every keyframe by hand,
one rig control at a time. The rig offers little guidance along
the way: its controls are numerical parameters whose values give
almost no hint of the pose they will produce, so the animator
cannot directly dial in the values for a desired look. Each
keyframe is instead found by trial and error: adjust a control,
render the view, compare against the intended look, and repeat
across many parameters until the pose reads correctly on screen.

We observe that, through every parameter tweak, the animator is
really steering toward a single contour: one that flows the way a
hand-drawn anime line would. The primacy of the line is not
incidental. 2D animation is fundamentally a line medium---every
frame begins and ends as a drawing whose information is
concentrated in its line work---and stylized 3D inherits that
aesthetic directly. Toon shading, the default for the style,
reinforces the point. Once the contour is right, the fine surface
detail inside it barely affects the final image, unlike
photorealistic rendering, where micro-geometry, shading, and
texture all drive the result. The contour is therefore both the
most economical signal we can ask an artist for and a sufficient
one to determine the look. This exposes an opportunity that no
current tool offers. If the animator could draw the contour
directly and have the rig conform, the entire trial-and-error
search would collapse into a single act of drawing. Closing that
gap is the \emph{central goal} of this paper: starting from an existing base pose,
we treat artist-redrawn contours over the rendered view as the control
input and recover the rig-parameter updates needed to stylize the pose to match them.
In this sense, our method can be viewed as a contour-driven form of
\emph{shot sculpting}.

\begin{figure}[b]
  \centering
  \includegraphics[width=\linewidth]{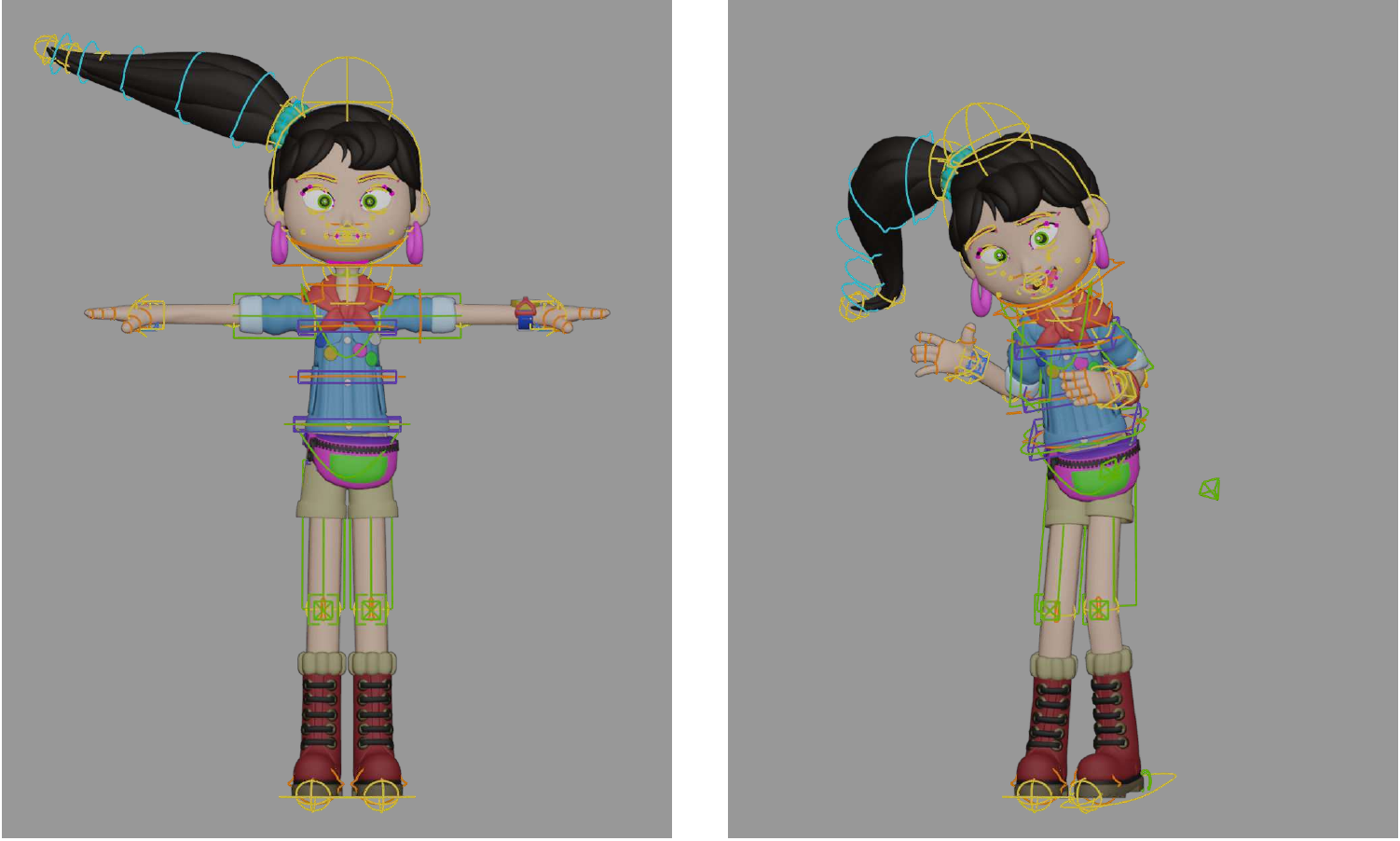}
  \caption{A production rig overlay on a stylized character.
  Left, the rest pose with the rig controllers visible.
  Right, a pose obtained by adjusting those controllers.
  Character model \textcopyright~Blender Studio.}
  \label{fig:rig}
\end{figure}

We recover the high-level rig controller parameters that an
animator works with in a DCC tool---the very layer at which
characters are posed in production. Optimizing at this layer is
what makes the result useful to the artist: the recovered pose
is a set of interpretable controller values rather than a baked
mesh, so it remains meaningful to the animator and fully
re-editable through the standard rig interface. \autoref{fig:rig}
shows what this layer looks like: a network of named
controllers---colored wireframe boxes, curves, and handles
overlaid on the mesh. An
animator authors a keyframe by adjusting the position,
rotation, and scale of those controllers; under the hood, the
rig drives the low-level deformers---skeletal skinning,
blendshapes, and cage- or curve-based deformers
(\autoref{sec:related-rigging})---translating each adjustment
into a deformation of the mesh.
The problem is that such a rig is not built for inversion. From
its parameters down to the vertices it is a dependency graph
evaluated strictly forward: it computes vertex positions from
parameter values but returns no gradients, and interleaves hard
constraints, discrete switches, and clamped ranges along the way.
To any external optimizer, it is a \emph{non-differentiable black
box}. We need a strategy that goes directly from the contour to
the artist-usable rig parameters.

To this end, we construct a fully differentiable pipeline from rig
parameters to a screen-space loss, so that the inversion can be driven
by backpropagation. At its core is a pre-trained \emph{NeuralRig}, a
differentiable surrogate of the production rig that maps rig parameters
to object-space vertex positions. Such surrogates were introduced to
accelerate the forward evaluation of film-quality rigs~\cite{BODO18,BODO20};
we instead exploit their differentiability, using the NeuralRig purely
as a gradient path for the inversion. At each iteration we project its
contour vertices to the image, match them against the user-drawn strokes,
and form a screen-space loss over the resulting correspondences;
backpropagating this loss through the projection and the NeuralRig
updates the rig parameters. The surrogate is never a substitute for the
rig: it supplies gradients during optimization only, and the recovered
parameters are written back into the DCC to drive the original production
rig.

We validate the method on cartoon-style and anime-style rigged
characters, covering facial expressions, rigged props,
view-dependent deformation and perspective exaggeration. Across
these settings, the method recovers rig parameters that
reproduce the artist's strokes within seconds per keyframe,
providing a line-drawing alternative to traditional
keyframe authoring.

Our contributions are:
\begin{itemize}
  \item \textbf{Direct, artist-friendly inversion of the
        production rig.} We invert straight to the high-level rig
        controllers an animator authors with, so every recovered
        pose is interpretable and re-editable through the standard
        rig interface.
  \item \textbf{A line-driven posing modality.} We design a
        screen-space optimization pipeline that takes strokes
        drawn over the rendered view as the per-keyframe input,
        matching the gesture a 2D animator already uses to specify
        a pose and aligning with the line-driven aesthetic of
        stylized 3D.
\item \textbf{A general formulation.} Our method asks only for a
      mesh and a set of parameters that drive its deformation;
      given these, it recovers the parameters from a drawn line,
      no matter how the rig works inside or what the mesh
      represents.
\end{itemize}

\section{Related Work}
\label{sec:related}

\subsection{Rigging Systems}
\label{sec:related-rigging}

A rig equips a character with a low-dimensional set of controls
that drive a deformation of its mesh, letting an animator pose
the character without touching vertices directly. Two deformation
primitives are ubiquitous: blendshapes, which express the
mesh as a linear combination of sculpted shape
targets~\cite{LARZ14}, and linear blend skinning, which
binds each vertex to a skeleton and blends the joint
transforms~\cite{MLT88}. Many other deformers can serve as a
rig's building blocks just as well, from free-form 
deformation~\cite{SP86} to curve-driven wire
deformers~\cite{SF98}.

Animation research seldom drives a production rig directly. It
works instead through generic parametric models, such as
3DMM~\cite{BV99} and FLAME~\cite{LBB17} for faces or
SMPL~\cite{LMR15} for bodies, that summarize a whole category of
shapes in a handful of parameters. Such models are general but
expressively limited, far simpler than a production rig used to author stylized animation.

\subsection{Rig Inversion}
\label{sec:related-inversion}

Recovering rig parameters from an observed signal, a problem generally
called rig inversion, has been
explored through several approaches. Lewis and Anjyo~\cite{LA10}
formulate direct blendshape manipulation as an inverse problem in semantic slider space.
Bickel et al.~\cite{BLB08} infer large-scale facial deformation from sparse
motion-capture markers and learn pose-space correctives for fine-scale wrinkles. Holden et
al.~\cite{HSK15} regress a character's joint configuration to rig parameters with Gaussian
processes, later comparing Gaussian-process and feedforward regressors~\cite{HSK17}.
Gustafson et al.~\cite{GLK20} accelerate iterative inversion with an offline-learned
analytic approximation of the rig's forward function. Marquis Bolduc and Phan~\cite{MP22}
learn a differentiable forward surrogate and a mesh-to-controls inverse. Omens et
al.~\cite{OTYF25} fine-tune facial-tracking rigs to produce semantically meaningful controls
while treating the tracker as a potentially non-differentiable black box.

\subsection{Differentiable Rig Surrogates}
\label{sec:related-surrogate}

Complex production rigs are expensive and non-differentiable, motivating
neural forward surrogates. Bailey et al. learn nonlinear residuals for
film-quality bodies~\cite{BODO18} and later approximate expressive facial
rigs and wrinkles~\cite{BODO20}. FaceBaker~\cite{RGM20} emphasizes fast,
portable facial rigs; Song et al.~\cite{SSR20} use differential coordinates
and a reconstruction subspace for fidelity; and Li et al.~\cite{LAH21}
learn skeletal articulation with neural corrective shapes.
These approaches primarily target efficient forward evaluation or
automatic rig construction.
Differentiability further supports learned rig inversion~\cite{MP22} and editable
facial retargeting~\cite{QSA23}. Our NeuralRig instead provides a gradient path
from screen-space contour loss to production-rig controls.

\subsection{Sketch-Based Editing and Posing}
\label{sec:related-sketch}

Sketch input has long served as an interface for 3D character control.
Motion Doodles~\cite{TBP04} maps cursive sketches to parameterized motion,
whereas Davis et al.~\cite{DAC06} reconstruct candidate 3D articulated
poses from drawings and let the user resolve ambiguity. Kraevoy et
al.~\cite{KSP09} alternate correspondence and deformation to fit a
template mesh to user-drawn occlusion contours. Hahn et
al.~\cite{HMC*15}
optimize arbitrary black-box rig parameters to align artist-designed,
rigged sketch abstractions with an input sketch.
Bessmeltsev et al. construct 3D character canvases from cartoon
drawings~\cite{BCV15} and pose rigged characters from gesture
drawings~\cite{BVS16}. For faces, DeepSketch2Face~\cite{HGY17} predicts a
parametric face model from coarse sketches, while Cetinaslan and
Orvalho~\cite{CO18} solve blendshape weights from strokes drawn on the
surface. Sketch2Pose~\cite{BB22} predicts perceptual pose cues and fits a
parametric body model to bitmap sketches, whereas
Sketch2PoseNet~\cite{WZW25} directly regresses human pose and shape; Unlu
et al.~\cite{USB22} combine learned inference with interactive mannequin
posing. Sketch2Anim~\cite{ZGX25} transfers storyboard sketches to motion
through estimated keyposes and trajectories. Squidgets~\cite{KCS25}
instead matches strokes to abstraction curves that act as
scene-manipulation widgets.
These methods span different production stages, from initial pose
generation to local scene editing.
Many of these methods are domain-specific or use strokes as pose cues
and manipulation handles rather than final-image curves. Hahn et
al.~\cite{HMC*15} are most closely related: their sketch abstractions drive
arbitrary black-box rigs using simplified or iconographic manipulation
handles that need not appear in the final rendered image. Our method
instead refines an existing pose through corrective strokes that redraw
visible contours, directly specifying the target screen-space appearance.

\section{Method}
\label{sec:method}

\subsection{Overview}
\label{sec:method-overview}

\begin{figure*}[!t]
  \centering
  \includegraphics[width=\linewidth]{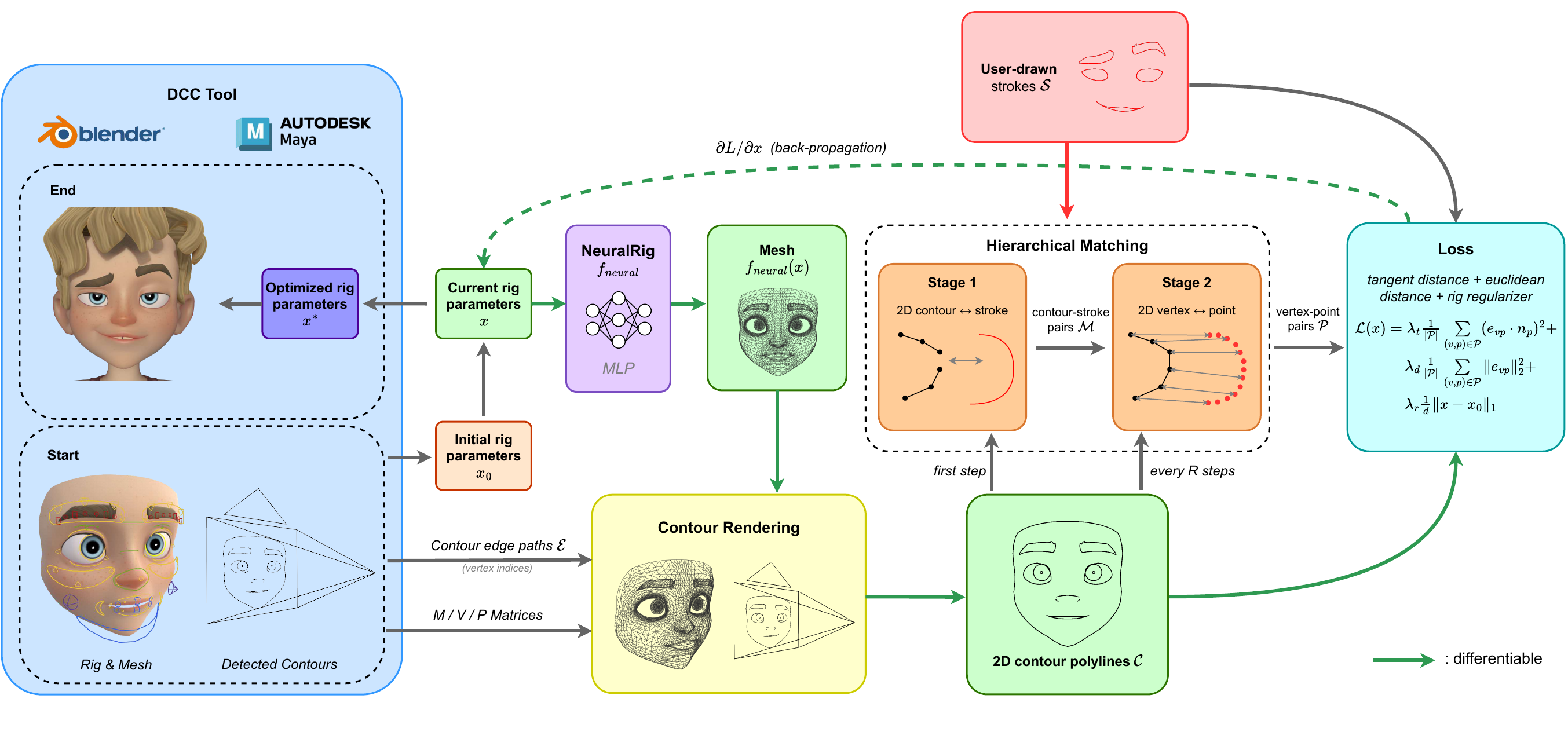}
  \caption{Pipeline overview. The DCC tool supplies an initial rig
  parameter vector $x_0$, the model and camera matrices $(M, V, P)$,
  and a set of fixed contour edge paths $\mathcal{E}$ along the visible
  contour of the current pose; the animator redraws those
  contours as strokes $\mathcal{S}$. The NeuralRig $f_{\text{neural}}$,
  a differentiable surrogate for the production rig, then maps the
  current rig parameters $x$ to mesh vertices; Contour Rendering
  filters by $\mathcal{E}$ and projects through $(M, V, P)$ to obtain
  2D contour polylines $\mathcal{C}$, which a two-stage matching pairs
  with $\mathcal{S}$ to produce vertex--point pairs $\mathcal{P}$.
  Backpropagating the screen-space loss over $\mathcal{P}$ updates
  $x$ at each Adam step, and the recovered $x^{\ast}$ is written
  back to the DCC as the new rig-parameter vector. Character model
  \textcopyright~Blender Studio.}
  \label{fig:pipeline}
\end{figure*}

Our pipeline (\autoref{fig:pipeline}) takes a rigged mesh at its
current pose, sitting in a DCC tool, such as Blender~\cite{Ble26}
or Maya~\cite{May26}. From this state the DCC provides
the current rig parameter vector $x_0$, the MVP
(Model--View--Projection) matrices $(M, V, P)$, and a set of
\emph{contour edge paths} $\mathcal{E} = \{E_j\}$---sequences of mesh-vertex
indices along the visible contour of the current pose.

While we describe $\mathcal{E}$ as contours throughout the paper,
the framework places no such requirement: any vertex sequence that
carries geometric information the optimization can drive the rig
against is valid. In practice, $\mathcal{E}$ can be obtained in several ways,
including an off-the-shelf line-drawing pipeline~\cite{BH19}, direct edge
selection, a screen-space selection drawn by the artist, or polylines authored
once ahead of time that bypass detection altogether. The paper does not commit to a single acquisition
route; how $\mathcal{E}$ enters the pipeline is an authoring
choice. In our Blender implementation, candidate contour edges are identified
from the opposite camera-facing signs of adjacent faces, with visible
boundaries and sharp creases optionally included. After detection, the user
can select the contours to edit and discard irrelevant candidates, yielding
the final $\mathcal{E}$ and narrowing the subsequent matching search for
greater accuracy and speed. The resulting $\mathcal{E}$ remains fixed
throughout optimization.

The animator then \emph{redraws} those $\mathcal{E}$ over the rendered view
as strokes $\mathcal{S} = \{S_i\}$; each stroke corresponds to an
existing contour edge path $E_j$ in screen space, possibly with shape
edits, rather than a freehand line introduced from scratch. Our method consumes
these inputs and recovers a rig parameter vector $x^{\ast}$ such
that the projection of $\mathcal{E}$ under the new pose agrees with
the strokes, then writes $x^{\ast}$ back into the DCC as the new
rig-parameter vector.

Formally, we seek
\begin{equation}
  x^{\ast} = \arg\min_{x}\,
  \mathcal{L}\!\left(
    \Pi_{M,V,P}\!\left(f_{\text{neural}}(x)\,\big|_{\mathcal{E}}\right),\;
    \mathcal{S}
  \right),
  \label{eq:problem}
\end{equation}
where $f_{\text{neural}}$ (the \emph{NeuralRig}) is a pre-trained differentiable surrogate for
the production rig,
$f_{\text{neural}}(x)\,\big|_{\mathcal{E}}$ restricts its object-space
vertex output to the indices in $\mathcal{E}$, and $\Pi_{M,V,P}$ lifts
those vertices to world space with $M$ and projects them through
$V, P$ to obtain the 2D contour polylines $\mathcal{C} = \{C_j\}$. The matrices
$M, V, P$, the contour edge paths $\mathcal{E}$, and the strokes $\mathcal{S}$
are all fixed during optimization; only $x$ is updated.

The optimization then runs as a loop. At each step the NeuralRig
maps the current $x$ to mesh vertices, and Contour Rendering
projects them to the 2D contours $\mathcal{C}$. The Matching
Process pairs $\mathcal{C}$ with the strokes $\mathcal{S}$ to
produce the correspondence $\mathcal{P}$. The screen-space loss
over $\mathcal{P}$ is then backpropagated through the projection
and the NeuralRig, and Adam updates $x$. The two matching stages
run on different schedules: the line-level pairing (Stage~1) is
computed once at the start and reused for the whole run, while
the point-level pairing (Stage~2) is refreshed periodically as
the mesh deforms, but not at every step, keeping the inner loop
cheap.

The remaining subsections describe each component in turn.
\autoref{sec:method-neuralrig} describes the NeuralRig, the
differentiable surrogate that maps the rig parameter vector to mesh
vertices. \autoref{sec:method-relevance} introduces the precomputed
\emph{rig influence map} $I$, used to restrict gradient
updates to the rig controls relevant to $\mathcal{E}$.
\autoref{sec:method-matching} details the two-stage hierarchical
matching that produces $\mathcal{P}$.
\autoref{sec:method-loss} defines the three loss terms and the
optimization loop.

\subsection{The Neural Rig}
\label{sec:method-neuralrig}

The NeuralRig $f_{\text{neural}}$ named in \autoref{eq:problem} is the
differentiable surrogate we use in place of the original production
rig. Production rigs are non-differentiable black boxes that
do not expose gradients to an external optimizer. 
$f_{\text{neural}}$ is an \emph{MLP}, trained once per
rigged mesh, that mimics the production rig's forward map from the rig
parameter vector $x$ to per-vertex positions in object space.

For production rigs, we narrow the NeuralRig along two axes to
keep training tractable. First, rather than training one NeuralRig
for the whole character, we train one per logical segment, such as
face, body, garment, or hair, so each segment-level NeuralRig only
has to fit its own subset of poses. Second, when $\mathcal{E}$ is known in advance for the segment, 
we further restrict the
NeuralRig's vertex set to the mesh patches that actually produce those
polylines, rather than the full closed mesh. In the face segment, for example,
we train only on the front-facing brow, eye, and face patches, so the NeuralRig
stays lightweight.

Collecting training data for the NeuralRig is a straightforward
task, because we already have a production rig.
For any rig parameter vector $x$ we choose, the DCC tool evaluates the production rig and
returns the resulting mesh. The components of $x$ are the
location, rotation, and scale parameters of the high-level rig
controls---the same handle values the animator authors on
keyframes and the system interpolates at playback. We
sample $x$ in two stages, first varying one rig control at a time
with the others held at rest, then varying several controls
simultaneously, so the dataset covers both single-control
responses and their interactions. 

For each sampled $x$, the per-vertex displacement that the DCC
returns---measured against the rest-pose mesh, i.e.\ the mesh
produced when every rig control sits at its identity value---serves
as ground truth; we write it $f_{\mathrm{GT}}(x)$. The NeuralRig is parameterized to
predict the same displacement, and we train it by regressing
$f_{\text{neural}}$ against $f_{\mathrm{GT}}$ on every vertex,
\begin{equation}
  \mathcal{L}_{\text{train}}
    = \mathbb{E}_{x \sim \mathcal{D}}
      \!\left[
        \frac{1}{V}\sum_{v=1}^{V}
          \bigl\lVert
            f_{\text{neural}}(x)_v - f_{\mathrm{GT}}(x)_v
          \bigr\rVert_{2}^{2}
      \right],
  \label{eq:train-loss}
\end{equation}
where $\mathcal{D}$ is the sampling distribution defined above and
$V$ is the rigged mesh's vertex count.

\subsection{Rig Influence Map}
\label{sec:method-relevance}

A naive backward pass through \autoref{eq:problem} updates every
rig control, including the many that have no effect on the
contour vertices in $\mathcal{E}$. We suppress those updates with
a precomputed binary \emph{rig influence map} $I$, where
$I_{k,v} = 1$ if rig control $k$ moves mesh vertex $v$ and
$I_{k,v} = 0$ otherwise. We build $I$ once per NeuralRig, before
optimization, by probing the production rig directly: at a few
base poses we perturb each control $k$ by $\pm\varepsilon$ and
set $I_{k,v} = 1$ whenever the resulting displacement of vertex
$v$ exceeds a small threshold.

From $I$ we keep only the controls that move the contour. These
form the active set
\begin{equation}
  \mathcal{A}
    = \bigl\{\,
        k \;:\;
        I_{k,v} = 1 \text{ for some } v \in \textstyle\bigcup_j E_j
      \,\bigr\}.
  \label{eq:active-set}
\end{equation}
At every gradient step we zero the gradient of every control
outside $\mathcal{A}$ and let Adam update only the rest. Since
$I$ and $\mathcal{E}$ are both fixed, $\mathcal{A}$ is computed
once and reused for the entire run.

\subsection{Hierarchical Matching}
\label{sec:method-matching}

\begin{figure*}[!t]
  \centering
  \includegraphics[width=\linewidth]{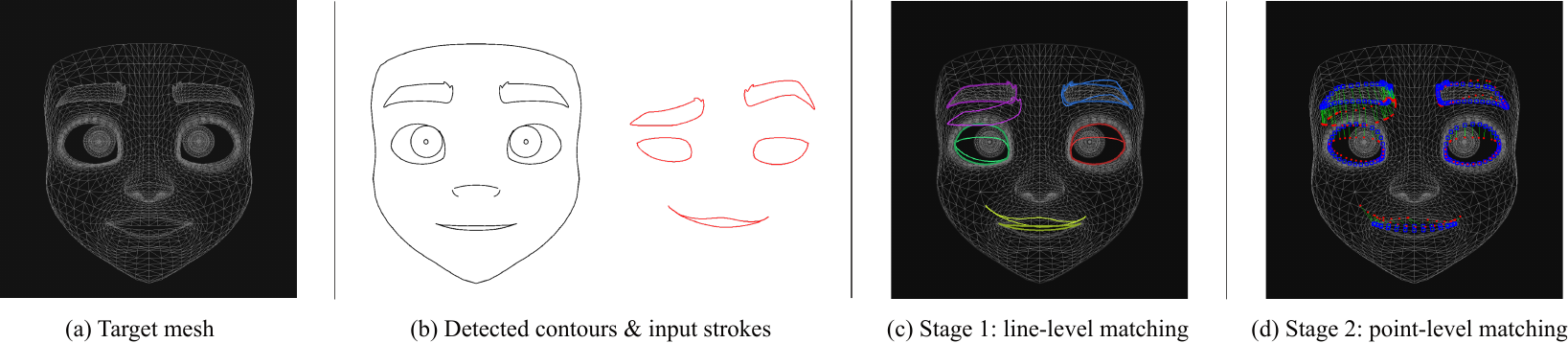}
  \caption{Two-stage hierarchical matching. (a)~The target mesh.
  (b)~Detected contours (black) and input strokes
  (red). (c)~Stage~1 pairs each input stroke with a contour; colors
  identify the line-level assignments. (d)~Stage~2 establishes ordered
  point-to-vertex correspondences within each assigned pair, which
  define the screen-space optimization loss. Character model
  \textcopyright~Blender Studio.}
  \label{fig:matching}
\end{figure*}

To evaluate the screen-space loss, we want to know which contour
vertex corresponds to which stroke point.
A flat assignment from every stroke point to every contour vertex
would scale combinatorially and ignore the natural topology of the
inputs. Strokes come as individual curves and the 2D contour
polylines $\mathcal{C}$ are ordered vertex sequences inherited from the underlying mesh,
so we factor the matching in two stages. Stage~1
(\autoref{sec:method-matching-stage1}) solves a coarse line-level
pairing between 2D contours and strokes, and Stage~2
(\autoref{sec:method-matching-stage2}) resolves the fine
vertex-to-point correspondence inside each pair.
\autoref{fig:matching} illustrates this two-stage procedure on
a facial example.

\subsubsection{Stage~1: Line-Level Matching}
\label{sec:method-matching-stage1}

Given the strokes $\mathcal{S}$ and the 2D contours $\mathcal{C}$
available at the initial matching step, we build a cost matrix $D$. Each entry
$D_{ij}$ measures how well stroke $S_i$ matches contour $C_j$ as a
curve. A stroke's points and a contour's vertices differ greatly in
density, so before computing the cost we resample both onto a
shared arc-length grid of $N$ points. Specifically, we use
$N=3(K_C-1)$, where $K_C$ is the number of vertices in the contour.
We write
$\tilde{S}, \tilde{C} \in \mathbb{R}^{N \times 2}$ for the two
resampled curves.

Open and closed curves are topologically different, so an open
curve should not correspond to a closed one; the matrix therefore
splits by the topology pair. Let $\chi(\cdot)$ denote a curve's topology
and let $\tilde{S}_i(\cdot)$ denote the
resampling of $S_i$ after applying the parametrization degree of
freedom that survives in each compatible case. The cost matrix is
\begin{equation}
  D_{ij}
    \;=\;
    \begin{cases}
      +\infty,
        & \chi(S_i) \neq \chi(C_j), \\[6pt]
      \displaystyle
      \min_{d \in \{\pm1\}}\,
        \mathrm{cost}\bigl(\tilde{S}_i(d),\, \tilde{C}_j\bigr),
        & \chi(S_i) = \chi(C_j) = \mathrm{open}, \\[10pt]
      \displaystyle
      \min_{o_S\in\mathbb{Z}_N}\,
        \mathrm{cost}\bigl(\tilde{S}_i(o_S),\, \tilde{C}_j\bigr),
        & \chi(S_i) = \chi(C_j) = \mathrm{closed}.
    \end{cases}
  \label{eq:stage1-cost-matrix}
\end{equation}

The two non-trivial branches differ only in the
parametrization freedom enumerated over $S_i$. For open curves,
the fixed endpoints leave only the traversal direction, so
$d\in\{+1,-1\}$ compares the native and reversed orders;
$\tilde{S}_i(d)$ denotes the corresponding resampling, and no
starting-point offset is required. For closed curves, we align
the screen-space windings by setting
$d^{\star}=\mathrm{sign}\!\bigl(A(S_i)A(C_j)\bigr)$, where
$A(\cdot)$ is the signed area, thereby fixing the direction before
matching. We then enumerate a cyclic offset
$o_S\in\mathbb{Z}_N$, where $\mathbb{Z}_N=\{0,\ldots,N-1\}$ denotes
the set of cyclic offsets, to rotate the starting point of
$S_i$ against the stored first vertex of $C_j$;
$\tilde{S}_i(o_S)$ denotes the resampling with orientation fixed
to $d^{\star}$ and its start rotated by $o_S$. Because the cost
depends only on the relative phase, an offset on one curve
suffices.

The shape cost itself takes two resampled curves and is the same
function in both branches. For inputs $\tilde{S}, \tilde{C}$, write
$L_S, L_C$ for their total arc lengths,
$\mathbf{x}_{S,n}, \mathbf{x}_{C,n} \in \mathbb{R}^{2}$ for their
$n$-th sample points, and
$\boldsymbol{\theta}_S, \boldsymbol{\theta}_C \in \mathbb{R}^{N}$ for
their \emph{turning functions}, where $\theta_{S,n}$ is the unwrapped
cumulative tangent angle of $\tilde{S}$ at the $n$-th sample, taken
with respect to a fixed screen-space reference direction. Because $\boldsymbol{\theta}$ is cumulative, an in-plane
rotation between two curves contributes a constant offset across all
$n$ in $\boldsymbol{\theta}_S - \boldsymbol{\theta}_C$. The cost combines three terms covering
shape, length, and per-point distance:
\begin{equation}
\begin{split}
  \mathrm{cost}(\tilde{S}, \tilde{C})
    \;=\;
    & w_{\theta}\, E_{\theta}(\tilde{S}, \tilde{C})
      \;+\; w_{L}\, \left| \tfrac{L_{S}}{L_{C}} - 1 \right| \\
    & \;+\; w_{d}\, \tfrac{1}{N} \sum_{n=1}^{N}
      \lVert \mathbf{x}_{S,n} - \mathbf{x}_{C,n} \rVert.
\end{split}
\label{eq:stage1-cost-terms}
\end{equation}
In our experiments, we use $w_{\theta}=1$, $w_{L}=0.5$, and
$w_{d}=1$.
The turning-function residual $E_{\theta}$ follows the metric of
Arkin et al.~\cite{ACH91}.
A closed-form alignment absorbs the global in-plane rotation
$\bar{\delta}$ between the two curves, leaving only the residual
turning-angle disagreement,
\begin{equation}
\begin{aligned}
  E_{\theta}(\tilde{S}, \tilde{C})
    &\;=\;
      \left[
        \tfrac{1}{N} \sum_{n=1}^{N}
        \bigl((\theta_{S,n} - \theta_{C,n}) - \bar{\delta}\bigr)^2
      \right]^{1/2}, \\
  \bar{\delta}
    &\;=\; \tfrac{1}{N} \sum_{n=1}^{N} (\theta_{S,n} - \theta_{C,n}).
\end{aligned}
\label{eq:stage1-turning-residual}
\end{equation}
We solve the resulting assignment problem on $D$ via the
Kuhn--Munkres algorithm~\cite{Kuh55,Mun57}, which
returns the assignment
\begin{equation}
  \varphi
    \;=\;
    \arg\min_{\varphi'}\,
    \sum_{i}\, D_{i,\,\varphi'(i)},
  \label{eq:stage1-assignment}
\end{equation}
with $\varphi'$ ranging over injections from strokes into contour
polylines, and collect the resulting line-level pairs into
\begin{equation}
  \mathcal{M}
    \;=\;
    \bigl\{\, (S_i,\, C_{\varphi(i)}) \,:\, i = 1, \dots, |\mathcal{S}|\, \bigr\}.
  \label{eq:stage1-pair-set}
\end{equation}
Stage~1 runs once at the first match and produces the pair set
$\mathcal{M}$. For each pair it also fixes the cost-minimizing
parametrization: the direction $d$ for open curves, or the cyclic
offset $o_S$ for closed ones. We apply it by reordering the
sample sequences stored in $\mathcal{S}$ and $\mathcal{C}$,
reversing their direction and rotating their start index so that
the paired $S_i$ and $C_{\varphi(i)}$ run the same way from a
corresponding origin. Stage~2 then operates on these aligned
pairs.

\subsubsection{Stage~2: Point-Level Matching}
\label{sec:method-matching-stage2}

For each pair $(S, C) \in \mathcal{M}$ with $K$ stroke points and
$L$ contour-polyline vertices, Stage~2 produces the vertex--point
pairs $(v, p) \in \mathcal{P}$ that the screen-space loss of
\autoref{sec:method-loss} consumes.

We build a $K \times L$ cost matrix $A$ whose entry $A_{ij}$
measures how compatible stroke point $i$ is with contour-polyline
vertex $j$. The cost is a weighted sum of three terms covering
arc-length position, tangent alignment, and screen-space
proximity,
\begin{equation}
\begin{aligned}
  A_{ij} = \;& \lambda_s\, (s_i - t_j)^2
              \;+\; \lambda_{\theta}\, \tfrac{1 -
                \langle \mathbf{t}^{S}_i,\, R(\bar{\delta})\,
                        \mathbf{t}^{C}_j \rangle}{2} \\
            +\;\;& \lambda_{a}\,
              \lVert \mathbf{p}_i - \mathbf{q}_j \rVert_2.
\end{aligned}
\label{eq:stage2-cost-matrix}
\end{equation}
The three terms encode complementary matching cues. The
arc-length term uses the normalized positions
$s_i,t_j\in[0,1]$ of stroke point $i$ and contour vertex $j$
along their respective curves, thereby preserving their relative
arc-length positions. The tangent term compares the unit tangents
$\mathbf{t}^{S}_i$ and $\mathbf{t}^{C}_j$ after the global
rotation is removed: $R(\bar{\delta})$ rotates the contour tangent
into the stroke frame, and the resulting half-versine
$\bigl(1-\langle\mathbf{t}^{S}_i,
R(\bar{\delta})\mathbf{t}^{C}_j\rangle\bigr)/2$
is zero for aligned tangents and lies in $[0,1]$. The Euclidean
term measures the screen-space distance between the pixel
positions $\mathbf{p}_i,\mathbf{q}_j\in\mathbb{R}^{2}$, favoring
the geometrically closer candidate when the other terms are
similar.

For the weighting parameters, we use a fixed tangent weight,
$\lambda_\theta=5$, while
$\lambda_s$ and $\lambda_a$ are length-adaptive. When $\widetilde S$ and
$\widetilde C$ have comparable screen length, the geometric
proximity term dominates; when one is much shorter than the other,
the geometric proximity becomes an unreliable signal and the
arc-length term takes over. Defining the screen-length ratio as
$r=\max(L_S,L_C)/\min(L_S,L_C)$, we set
$\lambda_s=20r$ and $\lambda_a=3/r$.

The matching is then found by a constrained \emph{dynamic time
warping} (DTW)~\cite{SC90} on $A$, a dynamic program with three
move types at each cell $(i, j)$,
\begin{equation}
  \Sigma_{ij}
    = \min\!\left\{
        \begin{aligned}
          & \Sigma_{i-1,\, j-1} + A_{ij}
              && \text{(match)}, \\
          & \Sigma_{i-1,\, j}   + \omega_i
              && \text{(skip point $i$)}, \\
          & \Sigma_{i,\, j-1}   + \infty
              && \text{(skip vertex $j$)},
        \end{aligned}
      \right.
  \label{eq:stage2-dp}
\end{equation}
where $\omega_i = 0$ for ordinary stroke points and
$\omega_i = +\infty$ for stroke corners. The disallowed vertex skip
forces every contour-polyline vertex to receive a match, and the
infinite corner-skip cost forces every detected stroke corner to
participate in the matching. Each diagonal step on the recovered
path emits one matched pair $(v, p)$ that joins $\mathcal{P}$.

As optimization deforms the mesh, its 2D contour $\mathcal{C}$
shifts and the current correspondence goes stale. We refresh
Stage~2 every $R$ iterations to track this. Each rematch rebuilds
the cost matrix $A$ from the up-to-date $\mathcal{C}$ and solves
the DTW anew.

\begin{figure*}[!t]
  \centering
  \includegraphics[width=\linewidth, height=0.9\textheight, keepaspectratio]{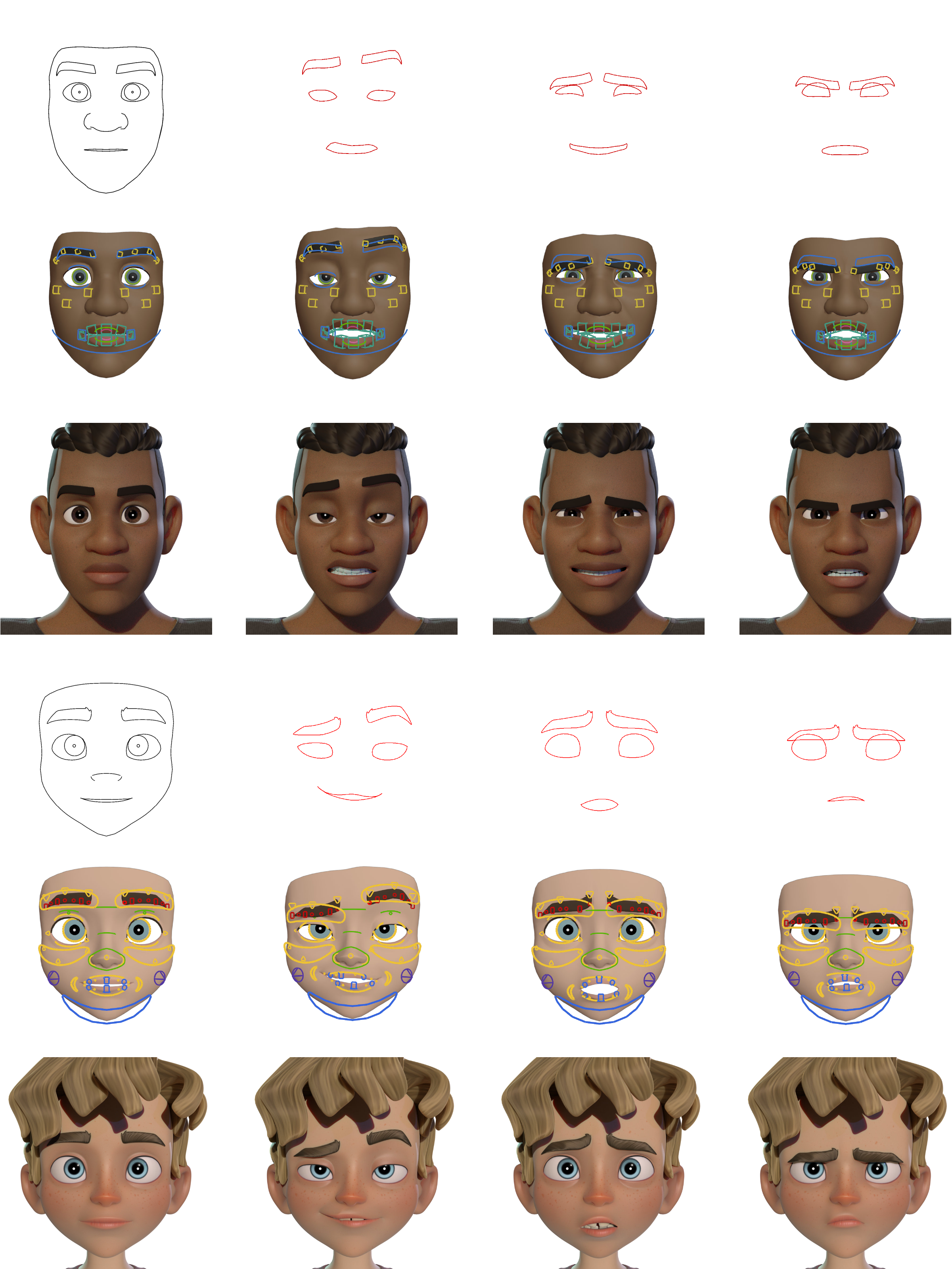}
  \caption{Facial expression keyframing on two cartoon-style
  rigged characters. Column~1 is the rest pose, included as
  reference; columns~2--4 are three recovered poses. For each
  pose, we show the input contour (top; rest-pose contour in
  black in column~1, artist's redrawn strokes in red in
  columns~2--4), the optimized mesh with the underlying rig
  controllers overlaid (middle), and the final rendered output
  (bottom). Character models \textcopyright~Blender Studio.}
  \label{fig:results-face}
\end{figure*}

\subsection{Loss and Optimization}
\label{sec:method-loss}

For each matched pair $(v, p) \in \mathcal{P}$ from
\autoref{sec:method-matching}, let $e_{vp} = \pi(v) - p$ denote
the 2D screen-space residual, where $\pi(\cdot)$ projects the
NeuralRig's object-space output for vertex $v$ through the matrices
$M$, $V$, $P$ onto the image plane. The objective combines two
weighted screen-space terms with a rig regularizer.

The dominant term penalizes only the component of $e_{vp}$ that
points across the stroke at $p$. Writing $n_p$ for the unit normal
of the stroke at $p$,
\begin{equation}
  \mathcal{L}_t
    = \frac{1}{|\mathcal{P}|}
      \sum_{(v, p) \in \mathcal{P}}
        \bigl(e_{vp} \cdot n_p\bigr)^2.
  \label{eq:loss-tangent}
\end{equation}
The tangent-direction component of $e_{vp}$ does not change the
visible contour, so projecting onto $n_p$ isolates the relevant normal
error.

When the contour is far from the strokes, the tangent term alone
provides an unstable update direction. The Euclidean term instead
pulls each vertex directly toward its matched point,
\begin{equation}
  \mathcal{L}_d
    = \frac{1}{|\mathcal{P}|}
      \sum_{(v, p) \in \mathcal{P}}
        \lVert e_{vp} \rVert_{2}^{2}.
  \label{eq:loss-euclidean}
\end{equation}
Rigs are often redundant, so multiple parameter settings can
produce similar screen-space results. The rig regularizer therefore
keeps the optimized parameter vector near the initial pose $x_0$,
\begin{equation}
  \mathcal{L}_r
    = \frac{1}{d}\, \lVert x - x_0 \rVert_{1},
  \label{eq:loss-rig}
\end{equation}
where $d$ is the dimensionality of $x$. We adapt $\lambda_r$
according to the relative reduction of the contour-fitting loss
$\mathcal{L}_{\mathrm{fit}}=\lambda_t\mathcal{L}_t+\lambda_d\mathcal{L}_d$,
increasing it as the contour-fitting loss decreases.

The full loss is the weighted sum
\begin{equation}
  \mathcal{L}
    = \lambda_t\, \mathcal{L}_t
      + \lambda_d\, \mathcal{L}_d
      + \lambda_r\, \mathcal{L}_r.
  \label{eq:loss-total}
\end{equation}
In our experiments, we use $\lambda_t=1$ and $\lambda_d=0.5$;
$\lambda_r$ increases from $1$ to $100$ during optimization as the contour-fitting loss decreases.
We minimize it with Adam using a fixed learning rate of $\eta=0.001$.
Before each Adam step, we zero gradients outside the active set
$\mathcal{A}$ (\autoref{sec:method-relevance}). Stage~2 is refreshed
every $R$ iterations, so the correspondences track the deforming mesh
without being recomputed at every step.

\section{Results}
\label{sec:results}

We evaluate the pipeline on stylized 3D characters. The four subsections that
follow exercise different facets of contour-driven rig inversion:
facial expression keyframing, contour-defined object posing,
view-specific shape stylization, and stylized perspective.
Each pose is recovered in a few to a dozen seconds. Matching
takes a negligible fraction of this; almost all of the runtime
is spent in the optimization loop, and it scales with the size
of the NeuralRig.

\subsection{Facial Expression Keyframing}
\label{sec:results-face}

Facial animation is a natural case for our method. Expressions
deform the mesh locally, so the depth ambiguity of 2D-line inversion
is mild. The face is also semantically dense: small changes around
the brows and mouth can distinguish expressions, allowing a few
cleanly recovered edits to span the range an animator would keyframe.

\autoref{fig:results-face} shows our method on two cartoon-style
face rigs. For each character, the
artist's strokes are the only per-pose input: each recovered
keyframe comes from a single optimization pass starting at the
rest pose. Across both characters and all three expressions, the rendered
contours of the recovered poses land on the user's strokes,
deforming the brows and mouth where the strokes specify movement
while leaving the eye and cheek regions close to the rest pose.
Because the output is a rig-parameter vector rather than a baked
mesh, each recovered keyframe remains editable through the
character's standard rig interface for further refinement by the animator.

\subsection{Contour-Defined Object Posing}
\label{sec:results-skirt}

\begin{figure}[H]
  \centering
  \includegraphics[width=\linewidth]{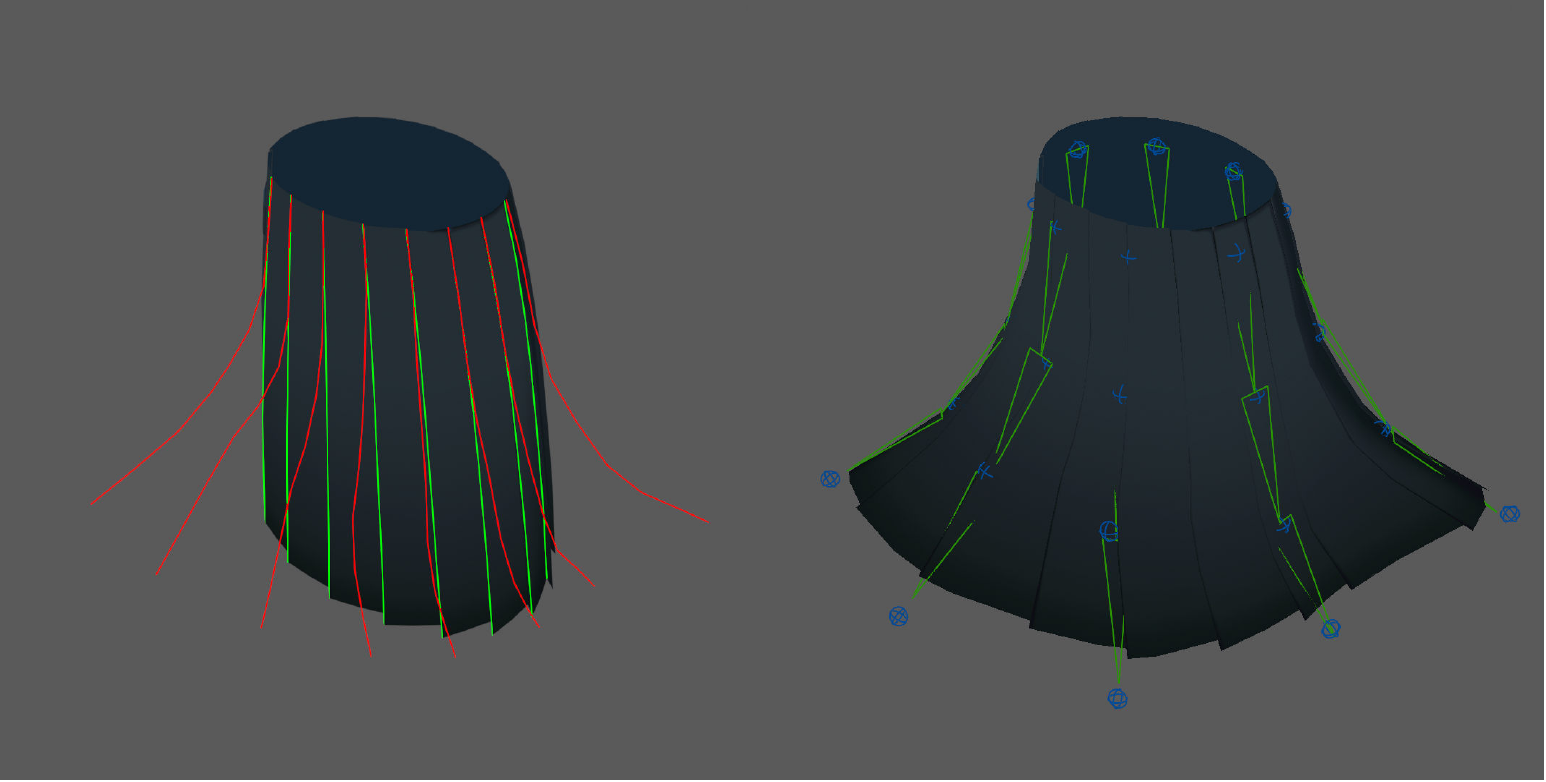}
  \caption{Posing a skirt rig through input strokes.
  Left, the rest pose: the projected rest-pose contour is shown
  in green and the artist's redrawn strokes are shown in red,
  asking the hem to flare outward. Right, the recovered pose,
  with the rig controllers overlaid. Character model
  \textcopyright~miHoYo.}
  \label{fig:results-skirt}
\end{figure}

Many objects
carry characteristic lines that encode the geometric
configuration of the object directly and serve as natural
posing handles. Redrawing these lines is a direct way an
animator can specify how the object should deform.
\autoref{fig:results-skirt} demonstrates this on a skirt rig.
This example also shows that our method handles poses requiring many
coordinated rig adjustments. Authoring the same pose by hand would
require locating and tuning each control against the rendered
view---a tedious search through hundreds of near-redundant degrees
of freedom. A single set of strokes instead conveys the posing
intent and drives all relevant controls in one optimization pass.

\subsection{View-Specific Shape Stylization}
\label{sec:results-view}

Stylized characters are designed for a camera: contour curvature
and landmarks tuned for one viewpoint may fail from another.
View-dependent deformation and animation have been studied~\cite{Rad99,CKB04,KI13,MA25},
but remain labor-intensive in production. Artists still correct
each offending camera angle by hand, whereas our method recovers
the desired correction from a single stroke drawn in the target view.

\begin{figure}[h]
  \centering
  \includegraphics[width=\linewidth]{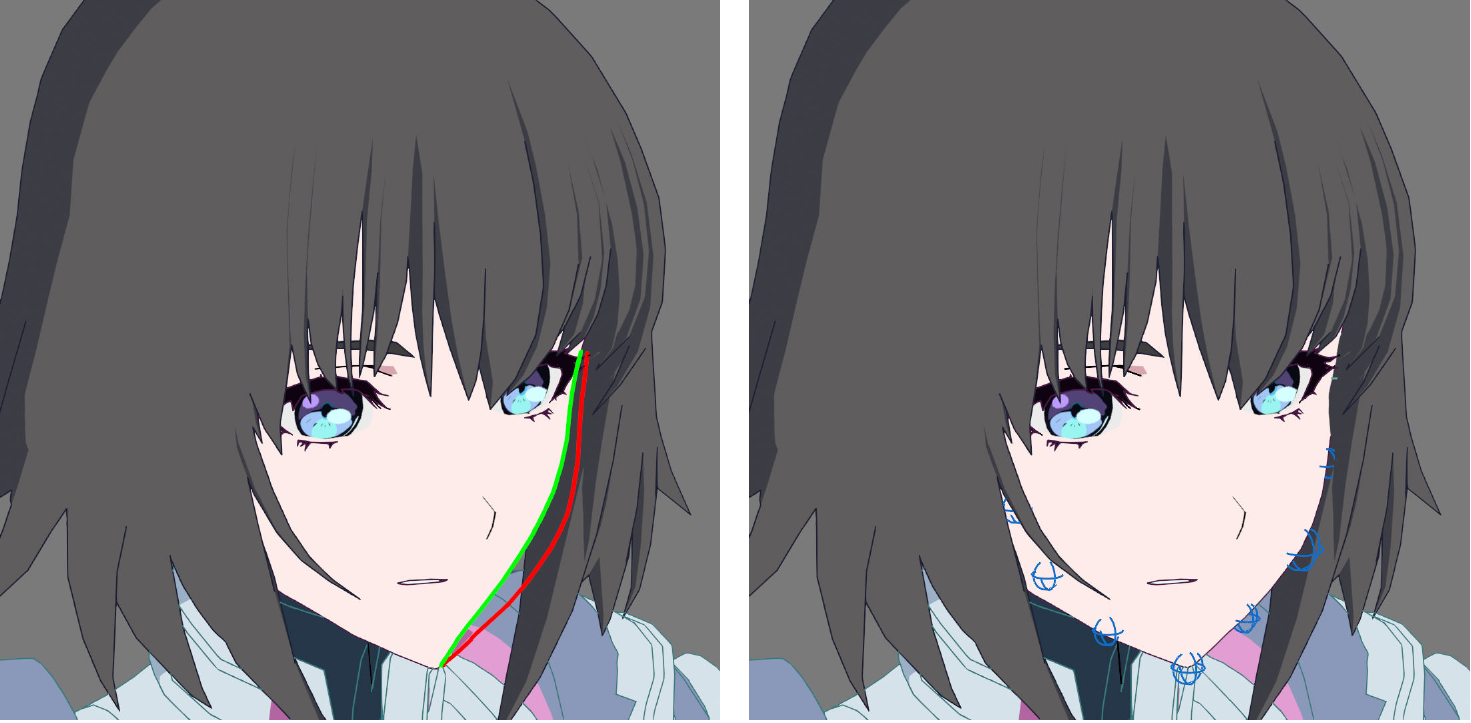}
  \caption{View-dependent contour adjustment on an anime-style face.
  Left, the rest pose: the rest-pose contour is shown in green
  and the artist's redrawn stroke is shown in red, asking for a
  different jaw-cheek curvature at this camera angle. Right,
  the recovered pose, with the rig controllers overlaid. Character
  model \textcopyright~DillonGoo Studios.}
  \label{fig:results-view}
\end{figure}

The most visible instance is the face contour of an anime
character. A 2D-illustrated face presents the jaw, cheek, and
chin as a continuous line whose curvature is tuned for the
camera; embodied as a 3D rig and rotated to a different angle,
that same line no longer reproduces the 2D-illustrated
curvature, and the face loses the cel-animation read.
\autoref{fig:results-view} shows what the method does with a
single stroke from the artist: the redrawn contour replaces the
current face contour, and the optimizer drives the rig until the
projected contour agrees.

\subsection{Stylized Perspective}
\label{sec:results-perspective}

Exaggerated perspective is another deliberate departure from
the world's geometry that stylized animation relies on for
visual impact. A 2D artist routinely breaks linear perspective
to make a punch reach further out of the page, an arm extend
dramatically toward the camera, or a sword loom larger than its
true depth would allow. Prior work has approached this problem
from the rendering side---Utsugi et al.~\cite{UNK11}, for
instance, build a multi-perspective control tree that warps the
rendered image to produce exaggerated illustration. In
production practice, however, the same effect is more commonly
delivered by reaching into the rig and reshaping the geometry
itself, once again at the cost of animator time.

Our method delivers this effect through the same stroke interface
used throughout. The artist redraws the contour of the part to be
exaggerated as it should appear under the intended projection, and
the optimizer recovers the rig parameters that produce it.
\autoref{fig:teaser}(c) shows the result on an anime-style
character: from the drawing camera, the initial pose reads flat,
but once the katana and the hand reaching toward the viewer are
redrawn beyond their true projected size, both loom toward the
camera---the perspective-breaking exaggeration a 2D artist
would draw, here carried by the character's own 3D geometry.

{\setlength{\columnsep}{8pt}\setlength{\intextsep}{2pt}%
\begin{wrapfigure}{r}{0.4\linewidth}
  \centering
  \includegraphics[width=\linewidth]{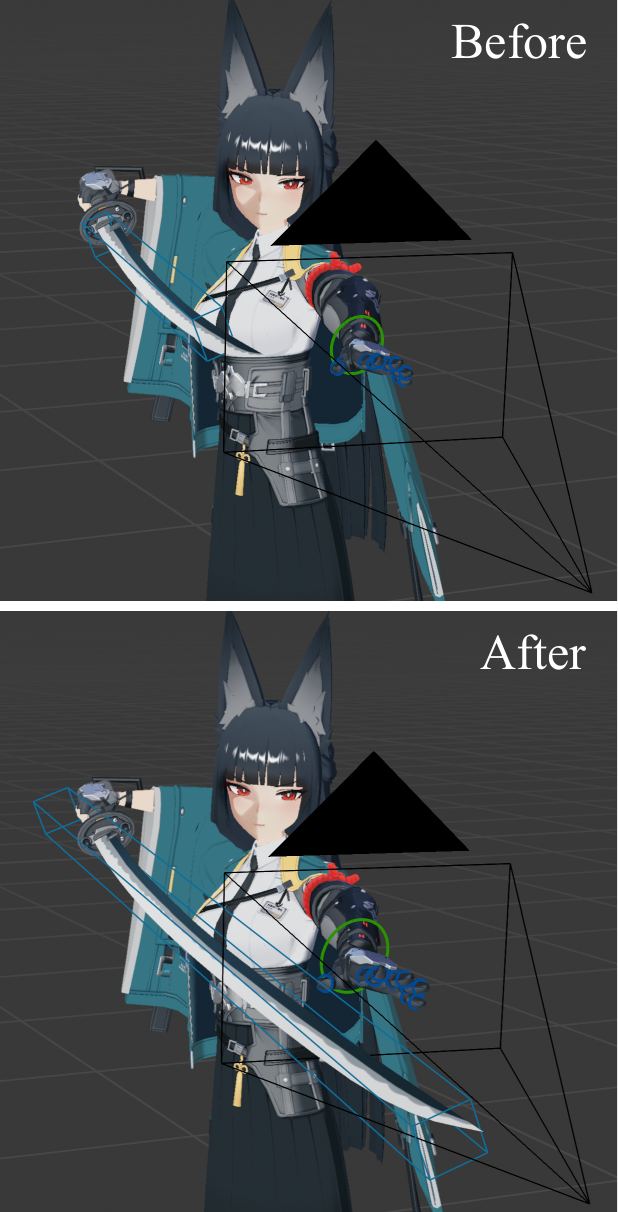}
\end{wrapfigure}
The inset at right provides an external view of the scene in
\autoref{fig:teaser}(c), before the edit (top) and after the edit (bottom). From this angle the recovered
geometry is openly distorted: the katana stretches far past any
natural length, sweeping across the body. That distortion is the
intent, not a failure. An exaggerated perspective only needs
to read correctly from its intended camera, so the geometry is shaped to look right
there and left to break from every other angle---a deliberate,
view-specific deformation that 3D artists otherwise dial in by
hand. Our method reaches the same result directly, from simple
strokes drawn on the camera view.\par}

\section{Discussion}
\label{sec:discussion}

\subsection{Limitations}
\label{sec:discussion-limits}

\paragraph*{Depth Ambiguity}
A 2D stroke carries no depth information. The method recovers
the artist's intent
reliably when the stroke commands only a local, subtle deformation---facial 
expression keyframing
(\autoref{sec:results-face}) is the clearest example. When the rig
admits enough deformation freedom, however, many rig parameter
configurations can lead to the same 2D contour; the
optimization still drives the loss down and lands the projected
contour on the artist's stroke, but the recovered pose 
is often not the one the artist had in mind. 
One possible solution is to let the stroke itself carry
depth information---for instance by mapping the brightness or saturation of the stroke color to relative depth.

\paragraph*{Fixed Contour Paths}
Our method keeps the vertex identities of the contour paths fixed
during optimization. This design is well suited to our target
scenario, namely shot-level refinement, in which the intended edits are
local and the initially visible contour generally remains within a stable
band of mesh edges. However, a larger deformation can change
visibility or contour topology: an initially selected path may disappear
while a different set of edges generates the rendered silhouette. The fixed
path can then align with the stroke even though the true silhouette does not.

\paragraph*{Surrogate Accuracy}
The recovered pose is only as accurate as the NeuralRig. Because
the NeuralRig is a learned approximation of the production rig,
the optimization minimizes the loss against the surrogate's
predicted contour rather than the real rig's, so any error in the
surrogate carries into the recovered parameters. This error is
largest for poses far from the training distribution, where the
surrogate is least reliable. The parameters we write back to the
DCC can therefore reproduce the artist's stroke slightly less
faithfully under the production rig than the optimization
indicated.

\subsection{Future Work}
\label{sec:discussion-future}

\paragraph*{Temporal Extension}
Extending the pipeline over time is one natural direction. The
pipeline currently solves a single keyframe at a time. If given
strokes drawn over a sequence of consecutive frames, the same
machinery could optimize the corresponding rig-parameter
trajectories jointly. We would add a smoothness loss on adjacent
frames' rig parameters, damping frame-to-frame jumps and yielding
a coherent animation rather than a sequence of independently
posed keyframes.

\paragraph*{Support for Secondary Motion}
Extending the method to secondary motion is an open problem. 
The NeuralRig is currently a static map from rig
parameters to vertex positions, but physics-driven secondary
motion---hair dynamics, cloth folds settling under gravity,
soft-body jiggle---is shaped by the trajectory the rig has
taken through pose space rather than by the current parameter
values. A static surrogate cannot represent this trajectory
dependence, so the method does not apply when the production
rig delegates part of its deformation to a simulator running in
the loop.

\paragraph*{Towards Real Production}
Production pipelines typically use several level-of-detail (LOD)
representations: animators block out poses on a coarse LOD and switch
to a detailed mesh for final refinement. Our method would therefore
require a separate NeuralRig for each LOD. Integrating this multi-LOD
workflow into a production-ready tool and evaluating it in a studio
pipeline remain future work.

\section{Conclusion}
\label{sec:conclusion}

We have presented a framework for shot-level stylization of rigged
characters through artist-drawn contour strokes. The artist
redraws an existing contour over the rendered view; a
pre-trained differentiable surrogate for the rig, paired with a
two-stage stroke-to-contour matcher and a screen-space loss,
propagates that edit back to the rig controls. We
demonstrated the framework on facial expression keyframing, object posing, view-specific shape stylization, and exaggerated
perspective on stylized characters. By recovering rig parameters directly from a redrawn contour,
our method gives animators a new, line-driven way to refine the
shot-specific appearance of stylized characters through the very lines the art form is built on.

\section*{Acknowledgements}
We used 3D character models credited to miHoYo, DillonGoo Studios,
and Blender Studio solely for research purposes. Some of the rigs
were obtained from the following publicly available sources:
\href{https://crabnuts.gumroad.com/l/crabmiyabirig}{Miyabi Rig},
\href{https://www.dillongoostudios.com/rigs}{DillonGoo Studios Rigs}, and
\href{https://studio.blender.org/characters/}{Blender Studio Characters}.
This work was supported by JST, CRONOS, Japan Grant Number JPMJCS25K1.

\bibliographystyle{eg-alpha-doi}
\bibliography{references}

\appendix
\section{NeuralRig Accuracy}
\label{sec:appendix-accuracy}

\begin{figure}[h]
  \centering
  \includegraphics[width=\linewidth]{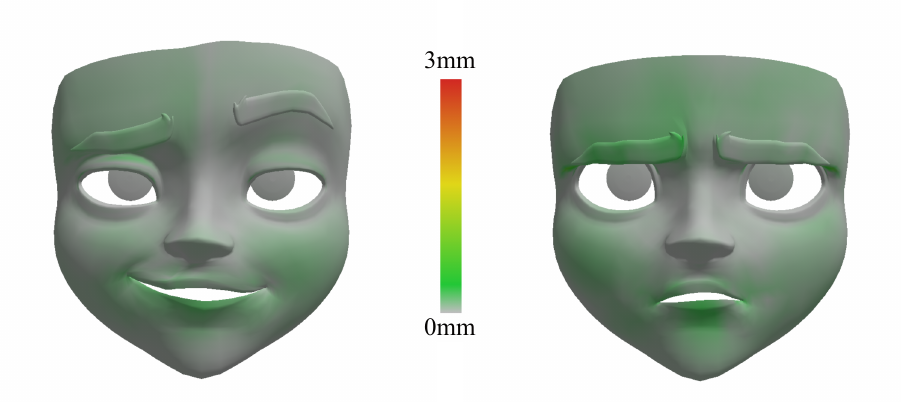}
  \caption{Per-vertex NeuralRig prediction error visualized as a
  surface heatmap.}
  \label{fig:neuralrig-error-heatmap}
\end{figure}

\begin{figure}[h]
  \centering
  \includegraphics[width=\linewidth]{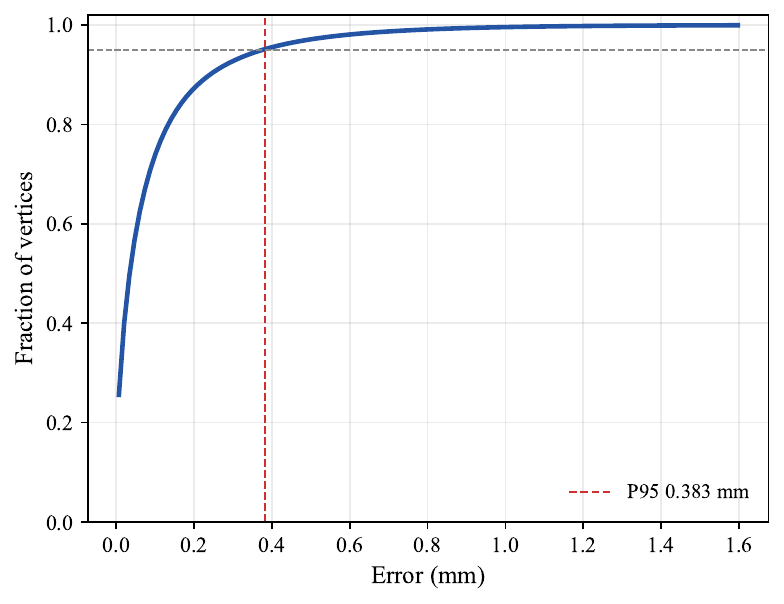}
  \caption{Cumulative distribution function of the per-vertex
  three-dimensional error, evaluated on 10,000 held-out
  poses.}
  \label{fig:neuralrig-error-cdf}
\end{figure}

We assess a single facial NeuralRig. \autoref{fig:neuralrig-error-heatmap}
visualizes its per-vertex three-dimensional error, while
\autoref{fig:neuralrig-error-cdf} reports the corresponding error distribution
over 10,000 held-out poses.

\section{Additional Experimental Details}
\label{sec:appendix-experiments}

\begin{figure*}[!t]
  \centering
  \includegraphics[width=0.98\textwidth]{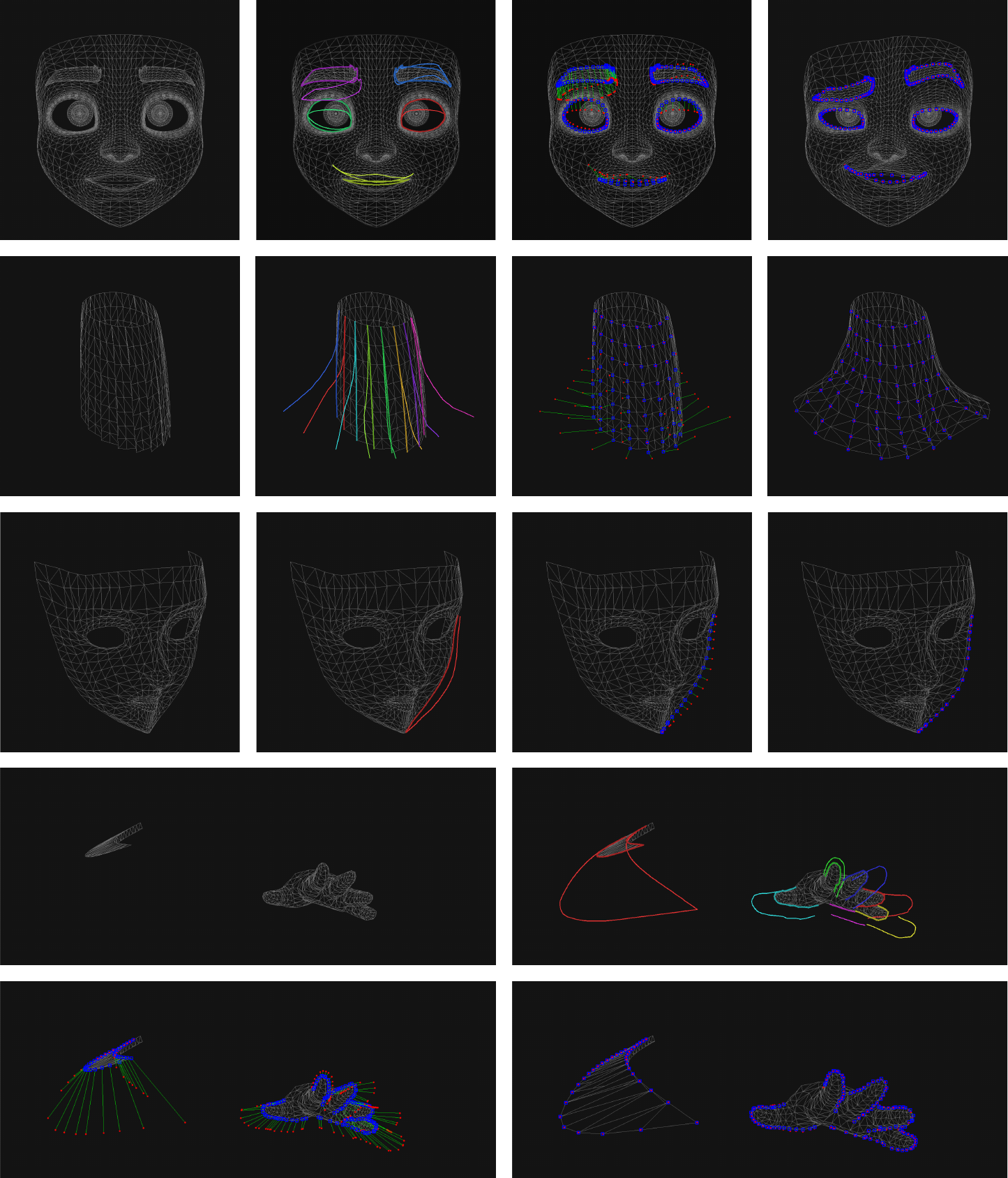}

  \caption{Additional results corresponding to the four applications in
  \autoref{sec:results}: facial-expression keyframing, contour-defined
  object posing, view-specific shape stylization, and stylized perspective.
  Models \textcopyright~Blender Studio, miHoYo, and DillonGoo Studios.}
  \label{fig:appendix-experiment}
\end{figure*}

\autoref{fig:appendix-experiment} presents representative results
for the four applications introduced in \autoref{sec:results}.
\autoref{tab:appendix-experiments} summarizes the corresponding NeuralRig
configurations, training settings, and optimization statistics.

\begin{table*}[t]
  \centering
  \small
  \setlength{\tabcolsep}{3.5pt}
  \renewcommand{\arraystretch}{1.25}
  \caption{Experimental details for the four experiments shown in
  \autoref{fig:appendix-experiment}. Model(s) lists
  the rigged asset(s); In/Out gives the NeuralRig input and output dimensions;
  MLP gives the number of hidden blocks and their width; Train. poses and
  Train. time report the training-set size and duration; Strokes and Vertices
  give the numbers of input strokes and matched contour vertices; and Steps
  and Time report the optimization iterations and total optimization time.
  Times were measured on an NVIDIA GeForce RTX 5070 Ti.}
  \label{tab:appendix-experiments}
  \begin{tabular}{l c c c c c c c c c}
    \toprule
    & \multicolumn{5}{c}{\textbf{NeuralRig}} &
      \multicolumn{4}{c}{\textbf{Optimization}} \\
    \cmidrule(lr){2-6}\cmidrule(lr){7-10}
    \textbf{Experiment} & \textbf{Model(s)} & \textbf{In/Out} &
    \textbf{MLP} & \textbf{Train. poses} & \textbf{Train. time} &
    \textbf{Strokes} & \textbf{Vertices} & \textbf{Steps} & \textbf{Time} \\
    \midrule
    Facial expression & face & 209/9720 &
      $4\times2048$ & 420,000 & 35~min & 5 & 448 & 240 & 6.3~s \\
    Object posing & skirt & 270/1476 &
      $4\times1024$ & 270,000 & 8~min & 8 & 72 & 300 & 13.3~s \\
    View-dependent correction & face & 81/2043 &
      $4\times1024$ & 40,000 & 3~min & 1 & 19 & 140 & 2.7~s \\
    \multirow{2}{*}{Perspective exaggeration} &
    katana & 8/171 & $4\times64$ & 20,000 & 2~min & 1 & 35 &
    \multirow{2}{*}{200} & \multirow{2}{*}{6.1~s} \\
    & hand & 93/1869 & $4\times1024$ & 280,000 & 4~min & 6 & 117 & & \\
    \bottomrule
  \end{tabular}
\end{table*}

\section{Extreme Input Strokes}
\label{sec:appendix-deviations}

\begin{figure}[h]
  \centering
  \includegraphics[width=\linewidth]{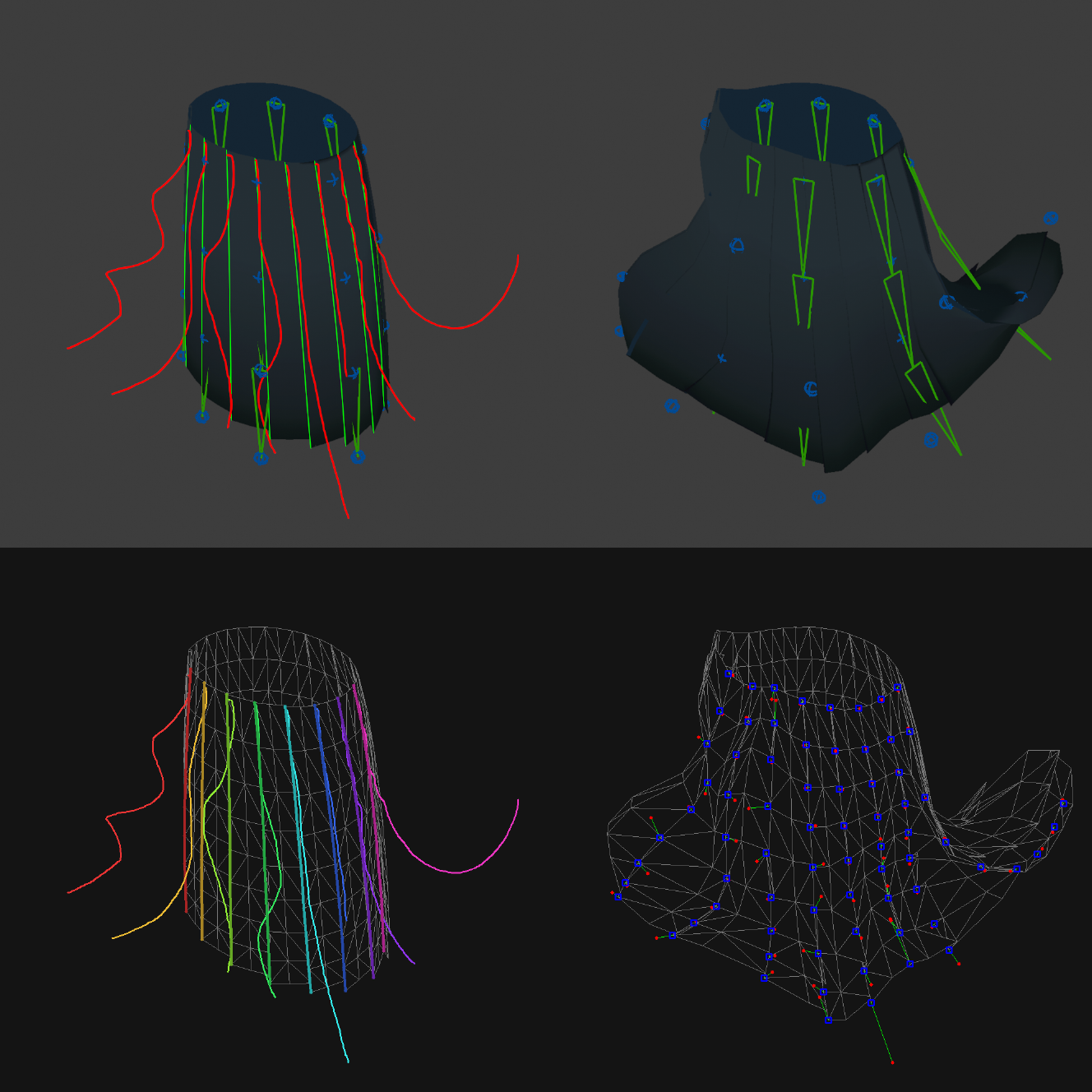}
  \caption{Optimization result for an extreme set of input strokes,
  illustrating a case outside the method's principal operating regime.}
  \label{fig:extreme-input-strokes}
\end{figure}

Substantial deviations between input strokes and existing contours do not
necessarily preclude matching: moderate differences remain tractable, and
ambiguous stroke-to-contour assignments can be specified by the user. The
harder problem is obtaining the intended deformation. Mesh topology, the
expressive capacity of the rig, depth ambiguity, and limitations of the
optimization can produce implausible 3D configurations even when the
projected contours align. Such cases lie outside our primary regime of
local, shot-level refinement. \autoref{fig:extreme-input-strokes}
illustrates a representative example.

\section{Freehand Input: Raw and Preprocessed Strokes}
\label{sec:appendix-preprocessing}

The experiments in the main paper use preprocessed strokes: we resample
them to approximately uniform spacing, suppress short spikes and high-frequency
jiggles, and optionally smooth the stroke while preserving detected corners.
Our intended users are experienced 2D artists, so their corrective strokes are
expected to be relatively clean.

\begin{figure}[h]
  \centering
  \includegraphics[width=\linewidth]{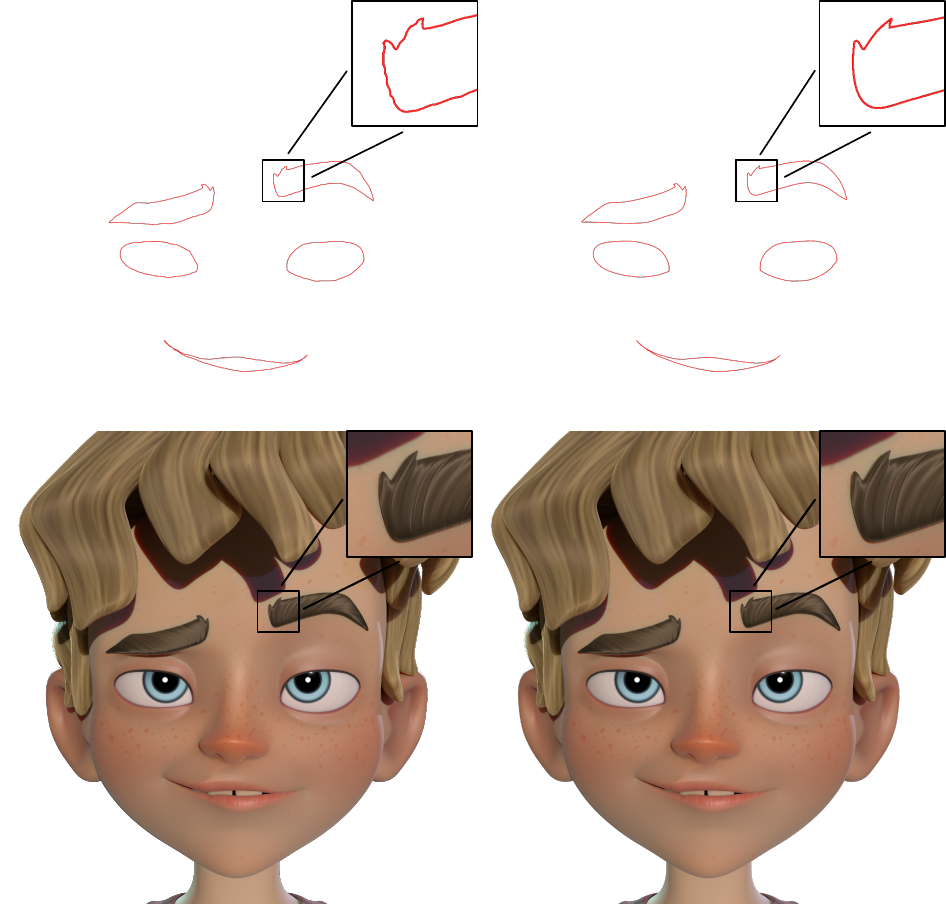}
  \caption{Comparison of optimization from raw freehand strokes (left)
  and preprocessed strokes (right). The difference is visually negligible
  in this tested example.}
  \label{fig:freehand-strokes-ablation}
\end{figure}

For comparison, we also optimize from raw,
unprocessed freehand strokes. \autoref{fig:freehand-strokes-ablation} shows
that the resulting difference is visually imperceptible. This is partly because the target
contours contain far fewer vertices than the input strokes, making the matching
primarily sensitive to the overall trajectory rather than point-level noise.
Small fluctuations therefore have limited effect, provided that they
do not create false corner detections during the point-level matching
described in \autoref{sec:method-matching-stage2}; such detections can
lead to incorrect constraints.

\end{document}